\documentclass[prl,twocolumn,showpacs]{revtex4}
\usepackage{graphicx}
\usepackage{graphics}
\usepackage{amsfonts}
\usepackage{amsmath}
\usepackage{epsfig}

\begin{document}
\catcode`\ä = \active \catcode`\ö = \active \catcode`\ü = \active
\catcode`\Ä = \active \catcode`\Ö = \active \catcode`\Ü = \active
\catcode`\ß = \active \catcode`\é = \active \catcode`\è = \active
\catcode`\ë = \active \catcode`\ô = \active \catcode`\ê = \active
\catcode`\ø = \active \catcode`\ò = \active \catcode`\í = \active
\catcode`\Ó = \active \catcode`\ú = \active \catcode`\á = \active
\catcode`\ã = \active
\defä{\"a} \defö{\"o} \defü{\"u} \defÄ{\"A} \defÖ{\"O} \defÜ{\"U} \defß{\ss} \defé{\'{e}}
\defè{\`{e}} \defë{\"{e}} \defô{\^{o}} \defê{\^{e}} \defø{\o} \defò{\`{o}} \defí{\'{i}}
\defÓ{\'{O}} \defú{\'{u}} \defá{\'{a}} \defã{\~{a}}



\newcommand{\li}{$^6$Li}
\newcommand{\na}{$^{23}$Na}
\newcommand{\cs}{$^{133}$Cs}
\newcommand{\kk}{$^{40}$K}
\newcommand{\rb}{$^{87}$Rb}
\newcommand{\vect}[1]{\mathbf #1}
\newcommand{\mf}{$m_F$}
\newcommand{\g}{g^{(2)}}
\newcommand{\one}{|1\rangle}
\newcommand{\two}{|2\rangle}
\newcommand{\limol}{$^6$Li$_2$}
\newcommand{\V}{V_{12}}
\newcommand{\kfa}{\frac{1}{k_F a}}

\title{Feshbach Resonances in Fermionic \li}

\author{C.H. Schunck, M.W. Zwierlein, C.A. Stan, S.M.F. Raupach, and W. Ketterle}

\affiliation{Department of Physics\mbox{,} MIT-Harvard Center for
Ultracold Atoms\mbox{,}
and Research Laboratory of Electronics,\\
MIT, Cambridge, MA 02139}

\author{A. Simoni, E. Tiesinga, C.J. Williams, and P.S. Julienne}
\affiliation{National Institute of Standards and Technology, 100
Bureau Drive stop 8423, Gaithersburg MD, 20899-8423, USA}

\date{\today}

\begin{abstract}
Feshbach resonances in \li\ were experimentally studied and
theoretically analyzed. In addition to two previously known
$s$-wave resonances, we found three $p$-wave resonances. Four of
these resonances are narrow and yield a precise value of the
singlet scattering length, but do not allow us to accurately
predict the location of the broad resonance near 83 mT. Its
position was previously measured in a molecule-dissociation
experiment for which we, here, discuss systematic shifts.
\end{abstract}

\maketitle

Interactions in ultracold atomic gases can be magnetically tuned
using Feshbach resonances. A Feshbach resonance occurs when the
energy of two colliding atoms is nearly degenerate to the energy
of a bound molecular state. Tunable interactions have been used to
explore novel phenomena in collisional and many-body physics.
Recently, Feshbach resonances have been used to control pairing
processes in ultracold fermionic gases. This led to the
observation of molecular Bose-Einstein condensates in \li\
\cite{joch03bec,bart04,zwie03molBEC,bour04coll} and \kk\
\cite{grei03mol_bec}, and to the first studies of the BEC-BCS
crossover, the continuous transition of fermion pairs from weakly
bound molecules to long range Cooper pairs
\cite{rega04,zwie04rescond,bart04,bour04coll,kina04sfluid,
bart04coll,chin04gap}.

In \li\ these experiments have been carried out in the vicinity of
the $s$-wave Feshbach resonance near 830 G
\cite{zwie04rescond,bart04,bour04coll,kina04sfluid,
bart04coll,chin04gap} (1 G $= 10^{-4}$ Tesla). The quantitative
interpretation of these experiments and the characterization of
the BEC-BCS crossover require a precise knowledge of the resonance
location. However, its determination is not trivial since the
resonance width is extremely large (180 G), and the line shape is
strongly affected by many body effects. In our previous work we
determined the position of this resonance by the onset of molecule
dissociation to be 822 $\pm$ 3 G \cite{zwie04rescond}.

In this paper we report on a detailed study of Feshbach resonances
in \li\ with the goal of accurately characterizing the interaction
potential of two \li\ atoms. Three resonances in the $|1\rangle$
and $|2\rangle$ states which are $p$-wave resonances have been
observed \cite{comment}. The positions of these Feshbach
resonances together with the location of a narrow $s$-wave
resonance in the $|1\rangle+|2\rangle$ mixture near 543 G are used
for a precise determination of the singlet $s$-wave scattering
length. These results, however, do not constrain the position of
the broad resonance, which also depends on the triplet scattering
length. An improved measurement of its location is presented and
the magnitude and the origin of possible systematic errors are
discussed.

The experimental setup has been described in Ref.\
\cite{hadz03big_fermi}. Up to $4 \times 10^7$ quantum degenerate
\li\ atoms in the $|F,$\mf$\rangle=|3/2, 3/2\rangle$ state were
obtained in a magnetic trap by sympathetic cooling with \na. The
\li\ atoms were then transferred into an optical dipole trap (ODT)
formed by a 1064 nm laser beam with a maximum power of 9 W. In the
optical trap three different samples were prepared: A single
radio-frequency sweep transferred the atoms to state $|1\rangle$
($|F, $\mf$ \rangle = |1/2, 1/2 \rangle$ at low field). Another
Landau-Zener sweep at an externally applied magnetic field of 565
G could then be used to either prepare the entire sample in state
$|2\rangle$ ($|1/2, -1/2 \rangle$ at low field) or create an equal
mixture of atoms in state $|1\rangle$ and $|2\rangle$. Except for
the measurement of the broad $s$-wave Feshbach resonance, all
resonances were observed by monitoring magnetic field dependent
atom losses. Atom numbers were obtained from absorption images
taken at zero field. The externally applied field was calibrated
by driving microwave transitions from state $|2\rangle$ to state
$|5\rangle$ ($|3/2, 1/2\rangle$ at low field) and from state
$|1\rangle$ to state $|6\rangle$ ($|3/2, 3/2\rangle$ at low field)
for several magnetic fields close to the resonance positions.

For spin polarized samples either in state $|1\rangle$ or
$|2\rangle$ $s$-wave scattering is forbidden by symmetry,
therefore the observed resonances occur in the $p$-wave channel.
The same molecular state that is responsible for these two
resonances also causes a $p$-wave resonance in the
$|1\rangle+|2\rangle$ mixture.

The three $p$-wave resonances were observed in clouds with typical
temperatures $T$ of 6 $\mu$K. This corresponds to $T/T_F$ in the
range of 0.5 to 1.5, where $T_F$ is the Fermi temperature. Radial
and axial trap frequencies were typically $\omega_r = 2 \pi \times
1.0$ kHz and $\omega_a = 2 \pi \times 6.9$ Hz.

\begin{figure}
\begin{center}
\includegraphics[width=3.5in]{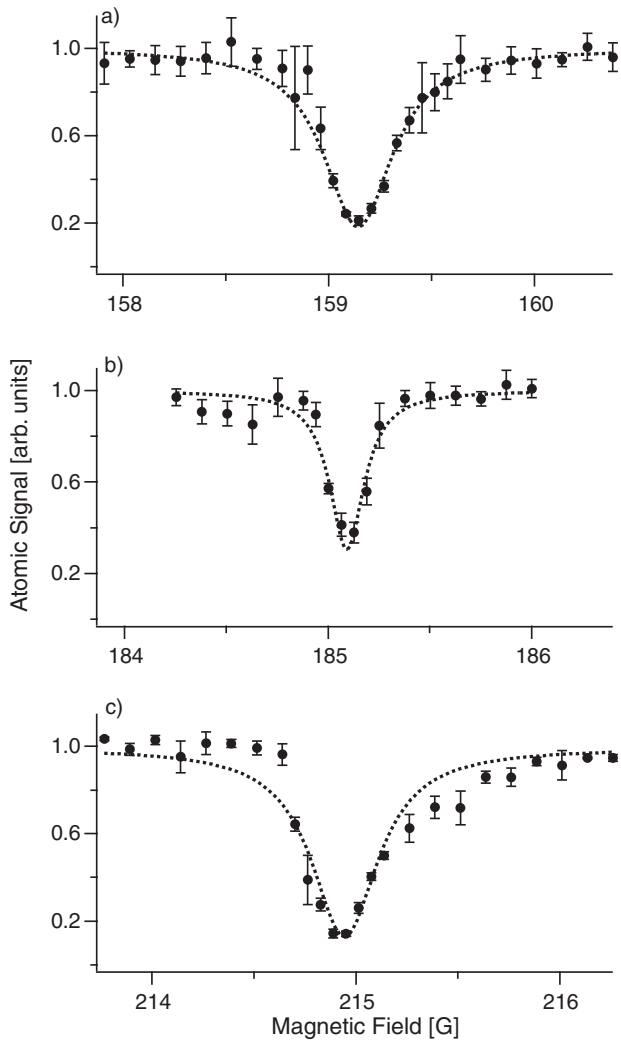}
\caption[Title]{$P$-wave resonances for $|1\rangle+|1\rangle$ (a),
$|1\rangle+|2\rangle$ (b), and $|2\rangle+|2\rangle$ (c)
collisions. The resonances were fitted by a Lorentzian. The
results are summarized in Table \ref{tab:resonancepos}.}
\label{fig:p-wave resonance1}
\end{center}
\end{figure}

The position of the p-wave resonance in the collision of a pair of
state $|1\rangle$ atoms was determined by first increasing the
magnetic field in 300 ms to approximately 5 G below the resonance.
Using an additional power supply to change the magnetic field
within a 10 G range, the field was then switched in 1 ms to a test
value $B_{\rm test}$. Here the atoms were kept for 200 ms before
the field and the optical trap were switched off. Finally atom
number versus $B_{\rm test}$ was recorded. Resonantly enhanced
losses due to inelastic three-body decay led to a Lorentzian
shaped feature as shown in Fig. \ref{fig:p-wave resonance1}(a).
Resonance positions and widths are summarized in Table
\ref{tab:resonancepos}.

The same technique was used to measure the $|1\rangle+|2\rangle$
and $|2\rangle+|2\rangle$, $p$-wave resonances. The observed
resonance lineshapes (see Fig. \ref{fig:p-wave resonance1}) are
non-symmetric, possibly due to losses while switching the magnetic
field to $B_{\rm test}$, or due to evaporative cooling which was
observed at magnetic fields $B_{\rm test}$ above the
$|1\rangle+|1\rangle$ and $|2\rangle+|2\rangle$ resonances.

The $s$-wave resonance near 543 G in the $|1\rangle+|2\rangle$
mixture was first observed in \cite{diec02fesh} and calculated in
\cite{ohar02}. Its position was determined as presented above in
clouds with typical temperatures of 6 $\mu$K, but in a slightly
deeper optical trap and with an extended holdtime of 2900 ms at
$B_{\rm test}$. The result of a fit to the Lorentzian lineshape is
given in Table \ref{tab:resonancepos}.

\begin{table}
\begin{ruledtabular}
\begin{tabular}{cccc}
States&$B_{\rm exp}$ [G]&$B_{\rm theory}$ [G]&Width [G]\\
\hline\\
$|1\rangle + |1\rangle$&$159.14 \pm 0.14$&159.15(4)&$0.4$\\
$|1\rangle + |2\rangle$&$185.09 \pm 0.08$&185.15(4)&$0.2$\\
$|2\rangle + |2\rangle$&$214.94 \pm 0.08$&214.90(4)&$0.4$\\
$|1\rangle + |2\rangle$&$543.28 \pm 0.08$&543.27(5)&$0.4$\\
$|1\rangle + |2\rangle$&$822...834$& \\
\end{tabular}
\end{ruledtabular}
\caption{\label{tab:resonancepos}Position of the Feshbach
resonances. Given are the experimentally and theoretically
determined resonance locations $B_{\rm exp}$ and $B_{\rm theory}$
respectively, and the measured resonance width. The uncertainties
for the experimental data in the first four rows are dominated by
magnetic field drifts between the measurement of the resonance and
the field calibration for which we measure an upper bound of 80
mG. For the $|1\rangle + |1\rangle$ resonance an additional drift
was monitored. The statistical error of determining the line
center, and the estimated uncertainty due to asymmetric line
shapes are negligible. For the broad $s$-wave resonance (fifth
row) only a range is given. See the text for a discussion.}
\end{table}

To determine the position of the broad Feshbach resonance near 830
G a different method was required. The resonance was identified as
the onset of molecular
dissociation~\cite{muka04,rega04,zwie04rescond}. Molecules were
first created on the repulsive (BEC) side of the Feshbach
resonance and then dissociated into atoms when the magnetic field
crossed the resonance.

\begin{figure}
\begin{center}
\includegraphics[width=3.5in]{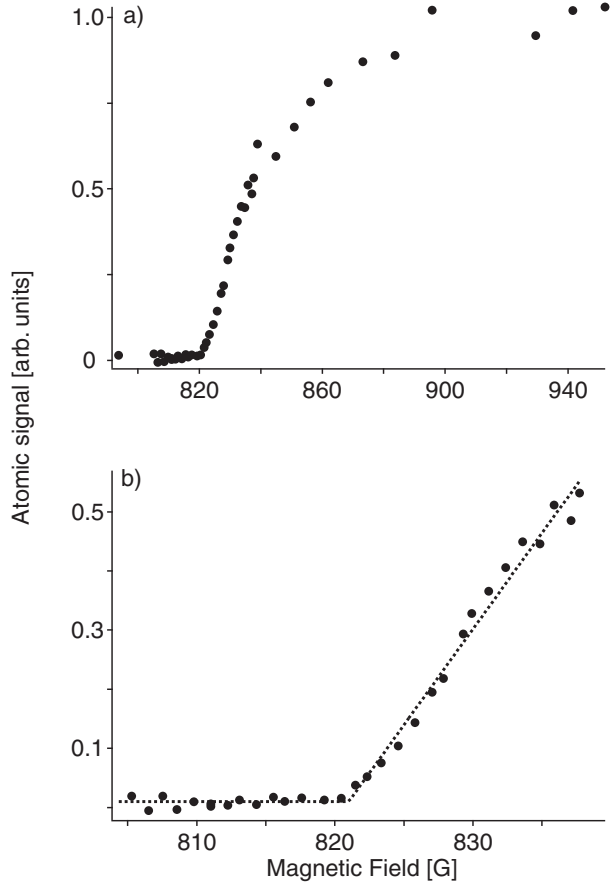}
\caption[Title]{Determination of the position of the broad
$s$-wave Feshbach resonance. (a) Onset of dissociation of
molecules into atoms at $821 \pm 1$ G. (b) The resonance position
was obtained by fitting two lines to the data points near the
threshold, one horizontal through the points showing no atomic
signal and a second line following the initial rise in atom
number. The crossing of those two lines gave the resonance
position, the estimated uncertainty in the crossing the quoted
error of $\pm 1$ G.} \label{fig:broad resonance}
\end{center}
\end{figure}

An almost pure \limol\ molecular BEC at a magnetic field of about
780 G was prepared in the optical trap as described in
Ref.~\cite{zwie04rescond}. The final ODT power was 36 mW, yielding
trap frequencies of $\omega = 2 \pi \times 690$ Hz radially and
$\omega = 2\pi \times 12.5$ Hz axially. The axial frequency has a
contribution from magnetic field curvature. After releasing the
molecules from the optical trap, the magnetic field was held at
780 G for 2 ms, before it was ramped to a test value $B_{\rm
test}$ in 14 ms. In these first 16 ms time of flight the molecular
peak density dropped to $n_{\rm mol} = 5 \times 10^9$ cm$^{-3}$.
The magnetic field was held at $B_{\rm test}$ for another 5 ms
before it was switched off in two steps: at a initial speed of 100
G/ms for 2 ms to leave the resonance region and then at a faster
speed (an exponential decay with initial time constant 30
G/$\mu$s) to zero field in 3 ms. Finally the sample was imaged
with light which was resonant only to atoms as the molecules are
detuned by about -1.3 GHz from the atomic transition at zero
field. By monitoring the atom number as a function of $B_{\rm
test}$ the onset of molecule dissociation was observed at $821 \pm
1$ G (Fig. \ref{fig:broad resonance}).

There are at least two sources of systematic error to be
considered. First, few-body collisions might dissociate molecules
when their size, which near resonance is on the order of the
scattering length between the constituent
atoms~\cite{kohl03feshbach}, becomes comparable to the mean
distance between the molecules. The scattering length near
resonance is parameterized by $a = a_{\rm bg}(1+\frac{\Delta
B}{(B-B_0)}) \approx a_{\rm bg} \frac{\Delta B}{(B-B_0)}$, where
$a_{\rm bg}$ is the negative background scattering length, $B_0$
is the resonance position, and $\Delta B$ is the resonance width.
Molecule dissociation will become important at a magnetic field
$B$ at which $a_{\rm bg} \frac{\Delta B}{(B-B_0)} \sim n_{\rm
mol}^{-\frac{1}{3}}$ and the scattering length is positive. For
our broad resonance this density-dependent few-body effect is
expected to shift the resonance position to lower magnetic fields.

The second systematic error is a density independent, single
molecule effect. Switching off the magnetic-field becomes
non-adiabatic close to resonance and destroys very-weakly bound
molecules~\cite{cubi03}. If $\dot{\omega}/{\omega}^2$, where
$\hbar \omega = \hbar^2/(ma^2)$ is the molecular binding energy
and $m$ is the atomic mass, becomes larger than unity, molecules
are forced to change their size too fast and may dissociate. This
systematically shifts the observed resonance position to lower
magnetic fields. The shift due to this ramp-induced dissociation
scales with $B-B_0 \sim {\dot{B}}^{\frac{1}{3}}$, where $\dot{B}$
is the rate at which the magnetic field is initially switched off
and $B$ determines the magnetic field where molecule dissociation
due to nonadiabaticity becomes important.

To determine the order of magnitude of these shifts we have
measured the resonance locations for three different ramp rates at
constant density and for three different densities at constant
ramp rate.

At a molecular density of $n_{\rm mol} = 1.5 \times 10^{10}$ the
resonance locations were measured at initial ramp speeds of 30
G/$\mu$s (fastest possible switch off), 100 G/ms (fastest
externally controlled ramp), and 12.5 G/ms (controlled ramp). For
the switch off the onset of dissociation occurs at $793 \pm 7$ G,
for the other two controlled ramps at $822 \pm 3$ G and no
relative shift is found within the errors. Assuming that no
density shifts affect these data, one can extrapolate to zero ramp
speed based on the $(B-B_0) \propto {\dot{B}}^{\frac{1}{3}}$
dependence. In this way we find a resonance position of $825 \pm
3$ G.

For a fixed initial ramp speed of 100 G/ms the resonance locations
were determined at densities of $5 \times 10^9$ cm$^{-3}$, $1.5
\times 10^{10}$ cm$^{-3}$ and $1.2 \times 10^{12}$ cm$^{-3}$ to be
$821 \pm 1$ G, $822 \pm 3$ G and $800 \pm 8$ G
respectively~\cite{density}. Here one can use the $(B-B_0) \propto
n^{1/3}$ dependence to extrapolate to a resonance position of $825
\pm 3$ G, neglecting effects due to nonadiabatic magnetic field
ramps.

Both systematic effects shift the maximum magnetic field value at
which the molecules are stable to lower magnetic fields. In a
simple picture, one would expect the total shift to be the larger
of the two. However, if they are similar, as in our case, they may
add up or combine in a more complicated way. We have measured the
threshold position at low density and slow ramp rates to be $822
\pm 3$ G and determined two shifts of $3 \pm 3$ G. Therefore, we
expect the position of the Feshbach resonance to be between 822
and 834 G. A more accurate extrapolation requires measuring the
dissociation threshold for more ramp speeds and densities.
However, technical limitations in varying magnetic field ramp
speeds and an unfavorable signal to noise ratio at lower densities
precluded this.

All Feshbach resonances discussed in this paper are due to the $v
= 38$ vibrational state of the singlet potential with total
electronic spin $S$ equal to zero. The $p$-wave resonances have a
total nuclear spin $I$ equal to one, while the 543 G and broad
$s$-wave resonances have $I=2$ and $I=0$, respectively.

The resonance locations are compared with results of scattering
coupled-channel calculations.  We locate the resonance from the
maximum of the elastic cross section as a function of magnetic
field. The collision energy is fixed at $E=k_BT$, where $k_B$ is
the Boltzmann constant and $T$ is the experimental temperature.
Our collision model, described in detail in Ref.~\cite{ohar02},
treats the singlet and triplet scattering length as adjustable
parameters. The triplet state has a total electron spin equal to
one. It turns out that all narrow resonances, which could be
accurately located, are insensitive to the triplet scattering
length. Only $s$ and $p$-waves are included in the calculation.
Fitting the singlet scattering length $a_S$ to the field locations
given in the first four rows of Table \ref{tab:resonancepos}
yields a very accurate value of $a_S=45.1591(16)~a_0$, where
$a_0=0.0529177$~nm. With this value, the resonance positions given
in the third column of Table~\ref{tab:resonancepos} were
calculated at a collision energy equal to $k_BT$. The agreement
with the experimental values is excellent. The $s$-wave resonance
is also in very good agreement with the determination of
Ref.~\cite{stre03}, 543.26(10)~G. Our theoretical uncertainties do
not include contributions due to a thermal average with respect to
the collision energy. Moreover, the shift between the field
values, at which the observed three-body loss rate and the
theoretical two-body elastic cross section are maximal, is
expected to be small but can not be ruled out at the current level
of accuracy.

The broad resonance is caused by a hyperfine-induced mixing
between a singlet vibrational level and an almost-bound virtual
state of the triplet potential, a situation analyzed
in~\cite{marc04, kemp04}. It is the virtual state that gives rise
to the large and negative triplet scattering length $a_T$ of \li.
Mixing occurs for magnetic field values above 500 G. In fact, in
absence of the hyperfine mixing the resonance would occur around
550 G. The coupling shifts the resonance by a few hundred Gauss.
For typical Feshbach resonances, these shifts are no more than a
few Gauss.  A consequence of the large shift is that the resonance
location depends critically on the less well known triplet
potential.

In conclusion, we have found three $p$-wave Feshbach resonances in
\li. They confirm the value of the singlet scattering length
determined from the narrow $s$-wave resonance of
Ref.~\cite{stre03}. The position of the broad resonances could not
be constrained using the refined singlet potential. The
determination of the position of the broad resonance via molecule
dissociation is subject to systematic errors, which shift the
onset of dissociation to lower magnetic fields.

The MIT research is supported by NSF, ONR, ARO, and NASA. S.
Raupach acknowledges financial support from the Dr. J\"urgen
Ulderup foundation.



\begin{thebibliography}{19}
\bibitem{density}
   \bibinfo{note}{The density was varied by using different parameters for
    the optical dipole trap, and by ramping the field to $B_{\rm
    test}$ either while the trap was on, or at different times of flight.}
\end{thebibliography}

\end{document}